**Work Design and Multidimensional AI Threat as**

**Predictors of Workplace AI Adoption and Depth of Use**

Aaron Reich[1], Diana Wolfe[2], Matt Price[3], Alice Choe[4], Fergus Kidd[5], Hannah Wagner[6]

[1] Chief Technology and Innovation Officer, Avanade

[2] PhD in Industrial Organizational Psychology; Vice President, Head of AI Research and Strategy, Kyndryl

[3] PhD in Media Psychology & Behavioral Neuropsychology; AI Governance, Responsible AI and Emerging Technology Platform Lead, Avanade

[4] PhD Candidate, Organizational Behavior and Human Resource Management, Rotman School of Management, University of Toronto

[5] Chief Technology Office, FieldPal.ai

[6] PhD Candidate, Industrial Organizational Psychology, Seattle Pacific University, 3307 3rd Ave. W, Seattle, WA

Correspondence concerning this article should be addressed to: aaron.reich@avanade.com

**Abstract**

Artificial intelligence (AI) tools are increasingly embedded in everyday work, yet employees' uptake varies widely even within the same organization. Drawing on sociotechnical and work design perspectives, this whitepaper examines whether motivational job characteristics (autonomy, skill variety, task significance, feedback) and multidimensional AI threat perceptions (changes in work, loss of controllability, loss of skills/expertise, loss of status/position) jointly predict workplace AI adoption and depth of use. Using cross-sectional survey data from employees ($N = 2,257$), we tested group differences across role level, years of experience, and region; multivariable predictors of AI adoption; and multivariable predictors of use depth, specifically frequency and duration. Across models, job design, especially skill variety and autonomy, showed the most consistent positive associations with AI adoption, whereas threat dimensions exhibited differentiated, sometimes mixed patterns for depth of use. Perceived changes in work was positively associated with frequency and duration, while status/position threat showed a directionally negative, but not consistently significant, relationship with deeper use. Findings are interpreted as associative, not causal, given the study design and reliance on self-report. Practical implications emphasize aligning AI enablement efforts with work design and monitoring potential workload expansion alongside adoption initiatives.

**Work Design and Multidimensional AI Threat as Predictors of Workplace AI Adoption**

**and Depth of Use**

Artificial intelligence (AI) tools are increasingly embedded in everyday work, but employees' uptake varies widely. Organizational behavior research argues that understanding AI at work requires moving beyond a "technology features" view to a sociotechnical view in which technology implementation reshapes work design, employee experiences, and downstream outcomes (Bankins et al., 2024; Parker & Grote, 2022). In parallel, research on digital transformation highlights that new technologies alter tasks, skills demands, and coordination, generating both opportunities for learning and concerns about the future of work (Trenerry et al., 2021).

This whitepaper integrates these perspectives by examining two sets of predictors of workplace AI use. First, we focus on core motivational job characteristics—autonomy, skill variety, task significance, and feedback—derived from the Job Characteristics Model and subsequent work design research (Hackman & Oldham, 1976; Morgeson & Humphrey, 2006). Second, we examine multidimensional AI threat perceptions, including perceived changes in work, perceived loss of controllability, perceived threats to skills and expertise, and perceived threats to status or position. Work on employee perceptions of smart technology and AI suggests that such appraisals are common and consequential for workplace attitudes and behavior (Brougham & Haar, 2018; Trenerry et al., 2021).

Bringing job design and AI threat together is practically important because organizations often try to increase AI usage through training and messaging alone. Work design theory suggests that structural features of jobs can either enable or constrain technology use by shaping opportunities to learn, experiment, and integrate tools into complex workflows (Parker & Grote, 2022). At the same time, research on AI–employee collaboration emphasizes that performance gains from human–AI collaboration depend on aligning technology with work systems and supporting employees' socialization into new ways of working (Chowdhury et al., 2022). We use this integrated lens to clarify which aspects of job design and which threat dimensions matter for AI adoption versus the depth of AI use.

**Theoretical Background**

***Job Design as Structural Enabler of Technology Adoption***

Work design provides a structural lens for understanding why some employees integrate AI tools into their work more than others. In the Job Characteristics Model (JCM), autonomy, skill variety, task significance, and feedback are theorized to promote internal work motivation and learning because they increase employees' experienced responsibility, meaningfulness, and knowledge of results (Hackman & Oldham, 1976). These job characteristics also anchor contemporary work design measurement, including the Work Design Questionnaire (WDQ; Morgeson & Humphrey, 2006), and meta-analytic evidence indicates that motivating job characteristics (including autonomy and task variety/complexity) reliably predict attitudes and performance-relevant outcomes (Humphrey et al., 2007). In addition, broader work design frameworks (e.g., job demands–resources theory) conceptualize autonomy, feedback, and skill use as job resources that help employees achieve work goals, cope with demands, and sustain learning (Bakker & Demerouti, 2007).

In a digital world, work design becomes even more consequential because technology can either enhance or erode the same job characteristics that drive engagement and learning. Sociotechnical perspectives emphasize that the outcomes of technology depend on how the technical system is embedded in the social system—that is, whether technology implementation supports joint optimization of work processes and human roles (Trist & Bamforth, 1951; Clegg, 2000; Parker & Grote, 2022). Consistent with this, Parker and Grote (2022) argue that digital technologies can shift job resources and demands in both directions: for example, AI-enabled systems can expand autonomy/control by distributing information and enabling localized decision making, but can also reduce autonomy by automating judgment and pushing workers "out of the loop" (Parker & Grote, 2022). They similarly argue that technology can increase "skill variety and use" by removing routine tasks and freeing workers to focus on more skilled, meaningful activities, yet it can also standardize work and reduce skill use through specialization and automation (Parker & Grote, 2022). Technology can likewise reshape feedback—sometimes improving access to performance information, but sometimes turning feedback into surveillance or punitive monitoring when embedded in algorithmic management systems (Parker & Grote, 2022).

These work design shifts matter for AI adoption and sustained use because job characteristics shape both (a) the opportunity structure for experimenting with AI, and (b) the ongoing fit between AI capabilities and the day-to-day demands of a role. Autonomy should facilitate initial experimentation because discretion gives employees latitude to trial new tools, adjust workflows, and decide how (and when) to incorporate AI into their tasks (Hackman & Oldham, 1976; Parker & Grote, 2022). Skill variety should facilitate deeper integration because roles that require diverse cognitive activities (e.g., drafting, synthesis, ideation, troubleshooting, and iterative refinement) create more recurring "use cases" for generative AI, increasing the likelihood of frequent and sustained usage (Morgeson & Humphrey, 2006; Parker & Grote, 2022). Feedback may support learning-by-using (e.g., employees can test outputs, evaluate quality, and refine prompts or workflows), yet feedback environments characterized by surveillance or deterrence can also discourage experimentation (Parker & Grote, 2022). Finally, task significance may create boundary conditions: where work has high stakes and accountability, employees may be more cautious about delegating or relying on AI outputs even if adoption occurs.

### AI Threat as a Multidimensional Psychological Barrier

Alongside structural enablers, employees' psychological responses to AI shape whether they embrace, avoid, or selectively engage with these tools. Research on "STARA" (smart technology, artificial intelligence, robotics, and algorithms) shows that employees form beliefs about whether advanced technologies could replace aspects of their work and what that implies for their longer-term career prospects (Brougham & Haar, 2018). Multilevel reviews similarly emphasize that employee responses to AI depend on what workers believe AI will do to their tasks, discretion, and opportunities for learning and development—and that these reactions are shaped by dynamics at multiple levels, including the nature of the technology, job roles, and organizational practices (Bankins et al., 2023).

Critically, AI threat perceptions are multidimensional rather than reducible to a single fear of displacement. Conceptually, employees can appraise AI as (a) intensifying change (e.g., disrupting workflows or altering task demands) and (b) producing perceived losses (e.g., reduced control, diminished expertise, or lower professional standing), and these appraisals can co-occur during technology implementation (Trenerry et al., 2021). Consistent with this view, research on AI-related identity processes argues that AI can threaten employees not only via job insecurity, but also via concerns about role value and status in the organization (Mirbabaie et al., 2021). In parallel, sociotechnical and work design scholarship highlights that new technologies can shift where control resides in the work system—changing how work is paced, monitored, and

coordinated—which can elicit concerns about reduced autonomy or controllability even when the technology is perceived as useful (Parker & Grote, 2022).

This multidimensional framing maps directly onto the threat dimensions examined here, which we derive from Mirbabaie et al.'s (2021) model of AI identity threat predictors. Specifically, we focus on four dimensions from their framework that capture both perceived AI-driven change and perceived losses. Changes in work reflects appraisals that AI is materially reshaping tasks, workflows, and expectations (Mirbabaie et al., 2021; Trenerry et al., 2021). Loss of controllability captures concerns that AI reduces discretion over how and when work is performed—an appraisal that work design theorists treat as central to job quality when technology changes the control–accountability balance in work systems (Mirbabaie et al., 2021; Parker & Grote, 2022). Loss of skills/expertise reflects concerns that AI may erode skill development or devalue specialized knowledge over time (Bankins et al., 2023; Mirbabaie et al., 2021; Trenerry et al., 2021). Loss of status/position captures identity-relevant concerns that roles like one's own may become less valued or less central, consistent with AI identity threat perspectives (Mirbabaie et al., 2021). Taken together, this literature supports treating AI threat as a set of separable appraisals that may relate differently to (a) whether employees try AI at all and (b) whether they integrate it deeply into everyday work (Bankins et al., 2023; Parker & Grote, 2022).

### Modeling Job Design and AI Threat Together

Sociotechnical perspectives suggest they must be modeled jointly to explain real-world adoption and sustained use (Clegg, 2000; Orlikowski, 2007; Trist & Bamforth, 1951; Parker & Grote, 2022). Work design shapes the opportunities employees have to incorporate AI into their jobs (e.g., discretion to experiment; task structures that create repeated use cases), while threat appraisals shape how employees interpret and respond to those same opportunities and constraints (Parker & Grote, 2022). In other words, even when employees view AI as potentially useful, their willingness to integrate it may hinge on whether it is perceived as undermining control, expertise, or role value (Mirbabaie et al., 2021; Trenerry et al., 2021).

This integrated approach aligns with multilevel evidence that AI outcomes in organizations are jointly shaped by technology characteristics, job roles, and organizational practices and supports (Bankins et al., 2023). It also aligns with research on work design in increasingly digital contexts, which argues that technology's effects depend on how it is embedded in work systems and how it changes the distribution of responsibility, control, and feedback (Parker & Grote, 2022). Finally, it complements human–AI collaboration perspectives emphasizing that performance gains depend on aligning AI capabilities with work processes and supporting employees as they develop routines for effective collaboration with AI (Chowdhury et al., 2022). In our framework, job characteristics capture the affordances and constraints of employees' work, whereas AI threat dimensions capture appraisals of what AI means for agency, competence, and professional standing. Modeling both together helps explain why some employees may adopt AI but use it shallowly, while others integrate AI into ongoing work routines (Bankins et al., 2023; Parker & Grote, 2022).

Bringing these perspectives together motivates a staged view of workplace AI engagement. From a work design standpoint, employees are more likely to *have viable opportunities* to try and integrate AI when their jobs include discretion and task complexity, conditions emphasized in classic and contemporary work design research as enabling experimentation and proactive changes in how work is done (Hackman & Oldham, 1976; Morgeson & Humphrey, 2006; Parker & Grote, 2022). From an appraisal standpoint, however, deeper integration may be shaped by whether employees interpret AI as expanding demands versus undermining control, expertise, or role value (Bankins et al., 2023; Mirbabaie et al., 2021; Trenerry et al., 2021).

Accordingly, in the present study we examine whether job characteristics (autonomy, skill variety, task significance, and feedback) and multidimensional AI threat perceptions (changes in work, loss of controllability, loss of skills/expertise, loss of status/position) explain (a) adoption (use vs. non-use) and (b) depth of use (frequency and duration). This combined model allows us to test whether structural job affordances account for meaningful variance above and beyond threat appraisals, and whether different threat dimensions show distinct associations with adoption versus depth—patterns that would be masked by treating "AI threat" as a single, undifferentiated construct (Bankins et al., 2023; Parker & Grote, 2022; Trenerry et al., 2021).

**Research Questions**

This study addresses three questions. First, how do job characteristics and AI threat vary across job level, years of experience, and region? Second, to what extent are job characteristics and AI threat associated with AI adoption and depth of AI use (frequency and duration)? Third, do job characteristics and AI threat predict AI use independently?

**Methods**

*Sample and Procedure*

Data were drawn from a cross-sectional organizational survey administered to employees across job levels, years of experience, and regions. Respondents reported job characteristics, perceptions of AI-related threats, and patterns of AI adoption and tool usage. The final analytic sample consisted of 2,257 respondents, though sample sizes vary across specific analyses due to missing data on particular measures. Because the design is cross-sectional, all findings are interpreted as associative rather than causal.

*Measures*

**Job characteristics (JCM).** Job characteristics were operationalized using four composites consistent with JCM constructs: autonomy, skill variety, task significance, and feedback (Hackman & Oldham, 1976; Morgeson & Humphrey, 2006). Autonomy represented the degree to which employees exercise discretion and control over work methods and scheduling. Skill variety captured the extent to which the job requires diverse skills and abilities. Task significance reflected perceptions that the work has meaningful impact on others. Feedback represented the degree to which employees receive clear information about performance effectiveness. Higher scores indicated more of the corresponding job characteristic (i.e., greater autonomy, greater skill variety, higher task significance, and more feedback). All items used 5-point Likert scales.

**AI threat.** AI threat was measured as four composites. *Changes in work* represented perceived increases in tasks and responsibilities resulting from AI. *Controllability* captured perceived reductions in control over structuring and executing work tasks. *Loss of skills and expertise* reflected concerns about the devaluation of current knowledge and competencies. *Loss of status and position* represented perceived threats to professional identity and organizational standing. Higher scores indicated greater perceived AI-related threat on each dimension (i.e., more perceived work changes, greater loss of controllability, greater perceived loss of skills/expertise, and greater perceived loss of status/position). All items used 5-point Likert scales.

**AI usage behaviors.** AI adoption was measured as a binary indicator of whether respondents reported using AI tools in their workflow (0 = no, 1 = yes). Frequency of use was measured on a six-point ordinal scale ranging from *0 = on rare occasions* to *5 = multiple times a day.* Duration of use captured the length of time respondents had incorporated AI into their workflow, rated on a six-point scale from *0 = less than one month* to *5 = over two years.* These items provided a basic behavioral profile of participants' AI adoption patterns.

### Analytic Strategy

Analyses proceeded in four steps. First, descriptive statistics characterized mean differences across job level, years of experience, and region. Second, ANOVAs tested whether job characteristics and AI threat dimensions varied significantly by demographic groupings, with effect sizes quantified using partial eta-squared ($\eta^2$) and false discovery rate (FDR) corrections applied to control Type I error (Benjamini & Hochberg, 1995). Third, multivariable regression models predicted AI adoption (logistic regression) and AI use frequency and duration (OLS regression), including all job characteristics and AI threat dimensions simultaneously. Variance inflation factors (VIF) were computed to assess multicollinearity, with values below 2.5 indicating acceptable levels (Hair et al., 2010; O'Brien, 2007).

**Results**

### Descriptive Patterns and Group Differences

**Job characteristics.** Job characteristics differed across job level, years of experience, and region (see Table 1 for descriptives; Table 2 for ANOVA tests). Autonomy and skill variety were generally higher at higher job levels. ANOVA results confirmed significant differences in autonomy across job levels, $F(5, 2222) = 24.19$, $p < .001$, $\eta^2 = .04$, and in skill variety, $F(5, 2227) = 31.21$, $p < .001$, $\eta^2 = .05$. Task significance and feedback did not vary significantly by job level ($p = .099$ and $p = .929$, respectively), suggesting that these dimensions may be more dependent on role design than hierarchical position.

Similar patterns emerged for years of experience, with autonomy, $F(5, 2224) = 13.11$, $p < .001$, $\eta^2 = .023$, and skill variety, $F(5, 2229) = 21.19$, $p < .001$, $\eta^2 = .037$, increasing with tenure. Regional differences also emerged for autonomy, $F(4, 2218) = 13.96$, $p < .001$, $\eta^2 = .019$, and skill variety, $F(4, 2223) = 13.54$, $p < .001$, $\eta^2 = .018$, reflecting possible cultural or organizational differences in work design across geographic locations (Hofstede, 2001; Leidner & Kayworth, 2006).

**Table 1**

*Job Characteristics Model Descriptives by Demographic Categories*

| | Category | n | Autonomy | Skill Variety | Task Significance | Feedback |
|---|---|---|---|---|---|---|
| **Role Level** | Analyst | 618 | 3.67 (0.71) | 3.92 (0.68) | 3.54 (0.63) | 3.70 (0.67) |
| | Consultant | 833 | 3.84 (0.68) | 4.07 (0.69) | 3.62 (0.60) | 3.69 (0.71) |
| | Director | 141 | 4.16 (0.51) | 4.46 (0.51) | 3.64 (0.64) | 3.74 (0.69) |
| | Manager | 641 | 3.98 (0.65) | 4.25 (0.62) | 3.62 (0.61) | 3.72 (0.71) |
| **Years of Experience** | <1 year | 141 | 3.68 (0.69) | 3.91 (0.62) | 3.44 (0.61) | 3.60 (0.62) |
| | 1–3 years | 365 | 3.71 (0.67) | 3.90 (0.65) | 3.55 (0.60) | 3.73 (0.65) |
| | 3–5 years | 310 | 3.76 (0.71) | 4.04 (0.72) | 3.57 (0.62) | 3.70 (0.76) |
| | 5–10 years | 436 | 3.86 (0.73) | 4.11 (0.68) | 3.61 (0.63) | 3.71 (0.72) |
| | 10+ years | 997 | 3.95 (0.65) | 4.23 (0.64) | 3.64 (0.61) | 3.71 (0.70) |
| **Region** | Europe | 1,010 | 3.86 (0.68) | 4.11 (0.69) | 3.55 (0.63) | 3.68 (0.69) |

| | | | | | |
|---|---|---|---|---|---|
| GDN | 37 | 3.77 (0.72) | 3.97 (0.77) | 3.78 (0.74) | 3.84 (0.71) |
| Growth Markets | 629 | 3.73 (0.71) | 4.00 (0.66) | 3.66 (0.59) | 3.69 (0.69) |
| North America | 560 | 3.99 (0.62) | 4.24 (0.61) | 3.61 (0.59) | 3.76 (0.71) |

*Note.* Values are *M* (*SD*) where M=mean and SD=standard deviation on a 5-point scale (1 = strongly disagree, 5 = strongly agree). Higher scores indicate *more* of the characteristic (i.e., greater autonomy, greater skill variety, greater task significance, and more feedback from the job).

**Table 2**

*ANOVA Results for Job Characteristics by Demographics*

| Demographic | Dependent Variable | F | df | p | η² |
|---|---|---|---|---|---|
| Role Level | Autonomy | 24.19 | 5, 2222 | <.001 | .042 |
| | Skill Variety | 31.21 | 5, 2227 | <.001 | .053 |
| | Task Significance | 1.95 | 5, 2234 | .099 | .003 |
| | Feedback | 0.22 | 5, 2234 | .929 | .000 |
| Years of Experience | Autonomy | 13.11 | 5, 2224 | <.001 | .023 |
| | Skill Variety | 21.19 | 5, 2229 | <.001 | .037 |
| | Task Significance | 4.46 | 5, 2236 | .001 | .008 |
| | Feedback | 1.03 | 5, 2236 | .001 | .008 |
| Region | Autonomy | 13.96 | 4, 2218 | <.001 | .019 |
| | Skill Variety | 13.54 | 4, 2223 | <.001 | .018 |
| | Task Significance | 5.07 | 4, 2230 | .002 | .007 |
| | Feedback | 1.95 | 4, 2230 | .119 | .003 |

*Note.* One-way ANOVAs tested mean differences in each job characteristic across demographic categories. Degrees of freedom are reported as (*df* between, *df* within). η² = eta squared effect size. p values are two-tailed.

**AI Threat.** AI threat perceptions varied by job level, experience, and region, but the pattern depended strongly on the threat dimension (see Table 3 for descriptives; Table 4 for ANOVA tests). Perceived changes in work, which refer to perceived increased tasks and responsibilities, varied significantly by region, $F(4, 2043) = 17.23$, $p < .001$, $\eta^2 = .025$, but showed weaker patterns by job level and experience. Controllability exhibited significant differences by job level, $F(5, 2045) = 5.26$, $p < .001$, $\eta^2 = .010$, with senior roles reporting lower perceived loss of control (Ng & Feldman, 2010). Status threat also varied by job level, $F(5, 2045) = 10.35$, $p < .001$, $\eta^2 = .020$, and by region, $F(4, 2042) = 4.81$, $p = .002$, $\eta^2 = .007$, suggesting that concerns about professional identity may be more pronounced at certain career stages and cultural contexts (Petriglieri, 2011).

ANOVA effect sizes were generally small ($\eta^2 < .06$). In particular, job level differences in autonomy and skill variety were among the largest effects in the descriptive set, underscoring that job characteristics remain meaningfully associated with hierarchical position even in contemporary knowledge work contexts.

**Table 3**

*AI Threat Perceptions by Demographic Categories*

| | Category | n | Work Changes | Controllability | Loss of Skills/ Expertise | Loss of Status/ Position |
|---|---|---|---|---|---|---|
| **Role Level** | Analyst | 562 | 3.19 (0.83) | 2.37 (0.70) | 2.76 (0.96) | 2.31 (0.88) |
| | Consultant | 752 | 3.21 (0.84) | 2.28 (0.69) | 2.74 (0.98) | 2.27 (0.89) |
| | Director | 133 | 3.29 (0.78) | 2.10 (0.63) | 2.47 (0.89) | 1.84 (0.79) |
| | Manager | 593 | 3.30 (0.81) | 2.24 (0.65) | 2.64 (0.91) | 2.13 (0.81) |
| | | | | | | |
| **Years of** | <1 year | 132 | 3.18 (0.71) | 2.30 (0.68) | 2.80 (0.94) | 2.38 (0.89) |
| **Experience** | 1–3 years | 329 | 3.17 (0.80) | 2.33 (0.67) | 2.79 (0.94) | 2.28 (0.89) |
| | 3–5 years | 286 | 3.19 (0.89) | 2.33 (0.70) | 2.70 (0.97) | 2.21 (0.85) |
| | 5–10 years | 396 | 3.20 (0.87) | 2.31 (0.72) | 2.69 (1.01) | 2.26 (0.91) |

| | | | | | |
|---|---|---|---|---|---|
| | 10+ years | 904 | 3.31 (0.80) | 2.23 (0.66) | 2.66 (0.92) | 2.14 (0.84) |
| **Region** | Europe | 899 | 3.10 (0.84) | 2.29 (0.67) | 2.65 (0.94) | 2.15 (0.82) |
| | GDN | 36 | 3.38 (0.84) | 2.27 (0.70) | 2.63 (1.12) | 2.24 (1.09) |
| | Growth Markets | 584 | 3.41 (0.75) | 2.33 (0.72) | 2.72 (0.98) | 2.32 (0.93) |
| | North America | 524 | 3.27 (0.83) | 2.22 (0.66) | 2.78 (0.93) | 2.20 (0.84) |

*Note.* Values are *M* (*SD*) where M=mean and SD=standard deviation. Higher scores indicate greater perceived threat for each dimension. *Controllability* items were reverse-coded so that higher scores reflect **greater loss of controllability** (i.e., higher threat).

**Table 4**

*ANOVA Results for AI Threat Perceptions by Demographics*

| Demographic | Dependent Variable | F | df | p | $\eta^2$ |
|---|---|---|---|---|---|
| **Role Level** | Work Changes | 1.77 | 5, 2046 | .133 | .003 |
| | Controllability | 5.26 | 5, 2045 | <.001 | .010 |
| | Skills/Expertise | 3.52 | 5, 2046 | .007 | .007 |
| | Status/Position | 10.35 | 5, 2045 | <.001 | .020 |
| **Years of Experience** | Work Changes | 2.87 | 5, 2047 | .022 | .006 |
| | Controllability | 2.03 | 5, 2046 | .088 | .004 |
| | Skills/Expertise | 1.52 | 5, 2047 | .193 | .003 |
| | Status/Position | 3.78 | 5, 2046 | .005 | .007 |
| **Region** | Work Changes | 17.23 | 4, 2043 | <.001 | .025 |
| | Controllability | 2.34 | 4, 2042 | .072 | .003 |
| | Skills/Expertise | 2.05 | 4, 2043 | .105 | .003 |
| | Status/Position | 4.81 | 4, 2042 | .002 | .007 |

**Note.** One-way ANOVAs tested mean differences in each job characteristic across demographic categories. Degrees of freedom are reported as (*df* between, *df* within). $\eta^2$ = eta squared effect size. p values are two-tailed.

*Multivariable Predictors of AI Adoption*

Logistic regression was conducted to predict AI adoption (0 = non-adopter, 1 = adopter) with all four job characteristics (autonomy, skill variety, task significance, and feedback) and all four AI threat dimensions (changes in work, controllability, skills and expertise, and status/position) entered simultaneously as predictors (see Table 5).

**Table 5**

*Logistic Regression Predicting AI Adoption*

| Predictor | β | SE | t | p | OR | 95% CI |
|---|---|---|---|---|---|---|
| **Job Characteristics** | | | | | | |
| Skill Variety | 0.31 | 0.06 | 5.64 | <.001 | 1.37 | [1.23, 1.52] |
| Autonomy | 0.20 | 0.06 | 3.32 | .001 | 1.22 | [1.08, 1.36] |
| Task Significance | −0.04 | 0.06 | −0.61 | .542 | 0.97 | [0.86, 1.08] |
| Feedback | −0.03 | 0.06 | −0.60 | .546 | 0.97 | [0.87, 1.08] |
| **AI Threat Perceptions** | | | | | | |
| Controllability | 0.11 | 0.05 | 2.01 | .044 | 1.12 | [1.00, 1.24] |
| Status/Position | −0.11 | 0.06 | −1.70 | .089 | 0.90 | [0.79, 1.02] |
| Skills/Expertise | 0.04 | 0.06 | 0.59 | .553 | 1.04 | [0.92, 1.18] |
| Work Changes | −0.02 | 0.05 | −0.40 | .686 | 0.98 | [0.89, 1.08] |

*Note.* n = 2,017. Model $\chi^2(8)$ = 77.56, p < .001, Pseudo $R^2$ = .029, Hosmer-Lemeshow $\chi^2(8)$ = 10.07, p = .260. OR = odds ratio; CI = confidence interval for OR.

The overall model was significant, $\chi^2(8)$ = 77.56, p < .001, McFadden's Pseudo $R^2$ = .029, indicating that the model with predictors was significantly better than the null model. The Hosmer-Lemeshow test moreover suggested adequate model fit, $\chi^2(8)$ = 10.07, p = .260. Variance inflation factors ranged from 1.07 to 1.91, indicating no multicollinearity concerns.

In logistic regression models predicting AI adoption, job characteristics emerged as central predictors. Skill variety showed the strongest positive association with adoption, β = 0.31,

*SE* = 0.06, *z* = 5.64, *p* < .001, OR = 1.37, 95% CI [1.23, 1.52], indicating that employees in roles requiring diverse competencies are significantly more likely to adopt AI tools. Autonomy also predicted adoption, β = 0.20, *SE* = 0.06, *z* = 3.32, *p* = .001, OR = 1.22, 95% CI [1.08, 1.36].

Loss of controllability exhibited a small positive association with adoption, β = 0.11, *SE* = 0.05, *z* = 2.01, *p* = .044, OR = 1.12, 95% CI [1.00, 1.24]. Task significance and feedback did not predict adoption (*p* = .542 and *p* = .546, respectively).

Other threat dimensions, including work changes and loss of skills and expertise, did not meaningfully predict adoption when job characteristics were included. This pattern suggests that although expertise-related concerns may exist, they do not independently influence the initial decision to adopt AI tools when structural job characteristics are accounted for.

### Multivariable Predictors of AI Use Frequency

A multiple regression analysis was conducted to predict AI use frequency (see Table 6). The model included the four job characteristics (autonomy, skill variety, task significance, and feedback) and four AI threat dimensions (changes in work, loss of controllability, loss of skills/expertise, and loss of status/position) entered simultaneously. The overall model was statistically significant, $F(8, 1209) = 4.48$, *p* < .001, $R^2 = .029$. Variance inflation factors ranged from 1.07 to 1.88, suggesting limited multicollinearity.

**Table 6**

*Multiple Regression Predicting AI Use Frequency*

| Predictor | β | SE | t | p | 95% CI |
|---|---|---|---|---|---|
| **Job Characteristics** | | | | | |
| Skill Variety | 0.15 | 0.04 | 4.14 | <.001 | [0.08, 0.22] |
| Feedback | −0.07 | 0.04 | −1.79 | .073 | [−0.14, 0.01] |
| Autonomy | −0.03 | 0.04 | −0.72 | .471 | [−0.10, 0.05] |
| Task Significance | 0.00 | 0.04 | 0.08 | .939 | [−0.07, 0.08] |

**AI Threat Perceptions**

| | | | | | |
|---|---|---|---|---|---|
| Work Changes | 0.06 | 0.03 | 1.97 | .049 | [0.00, 0.13] |
| Controllability | −0.06 | 0.04 | −1.70 | .090 | [−0.13, 0.01] |
| Status/Position | −0.06 | 0.04 | −1.47 | .142 | [−0.15, 0.02] |
| Skills/Expertise | 0.00 | 0.04 | 0.03 | .974 | [−0.08, 0.08] |

*Note.* $n = 1{,}218$. The overall model was statistically significant, $F(8, 1209) = 4.48$, $p < .001$, $R^2 = .029$, adjusted $R^2 = .022$. $\beta$ = standardized regression coefficient; the reported SEs and 95% CIs correspond to the standardized coefficients. CI = confidence interval.

Skill variety was a positive predictor of AI use frequency ($\beta = 0.15$, SE = 0.04, t = 4.13, p < .001, 95% CI [0.08, 0.22]), suggesting that AI tools may be used more often in roles characterized by more diverse and complex task demands. Changes in work was also positively associated with frequency ($\beta = 0.06$, SE = 0.03, t = 1.97, p = .049, 95% CI [0.00, 0.13]), consistent with the possibility that perceived work intensification covaries with more frequent AI use (Green, 2004; Kelliher & Anderson, 2010). Feedback was negatively associated with frequency but did not reach conventional significance ($\beta = −0.07$, SE = 0.04, t = −1.79, p = .073, 95% CI [−0.14, 0.01]). Loss of controllability and loss of status/position were directionally negative but not statistically significant in the model (both *p*s > .05).

### *Multivariable Predictors of AI Use Duration*

Multiple regression analysis was conducted to predict AI use duration (see Table 7). All four job characteristics (autonomy, skill variety, task significance, and feedback) and all four AI threat dimensions (changes in work, controllability, skills and expertise, and status/position) were entered simultaneously as predictors. The overall model was significant, $F(8, 1210) = 6.25$, p < .001, $R^2 = .040$, adjusted $R^2 = .033$. VIF values ranged from 1.07 to 1.92, indicating acceptable levels of multicollinearity.

**Table 7**

*Multiple Regression Predicting AI Use Duration*

| Predictor | β | SE | t | p | 95% CI |
|---|---|---|---|---|---|
| **Job Characteristics** | | | | | |
| Skill Variety | 0.13 | 0.04 | 3.22 | .001 | [0.05, 0.21] |
| Task Significance | −0.10 | 0.04 | −2.29 | .022 | [−0.18, −0.01] |
| Autonomy | 0.03 | 0.04 | 0.73 | .468 | [−0.05, 0.12] |
| Feedback | 0.00 | 0.04 | −0.05 | .960 | [−0.08, 0.08] |
| **AI Threat Perceptions** | | | | | |
| Work Changes | 0.17 | 0.04 | 4.92 | <.001 | [0.10, 0.24] |
| Status/Position | −0.09 | 0.05 | −1.87 | .062 | [−0.18, 0.00] |
| Skills/Expertise | −0.02 | 0.05 | −0.42 | .672 | [−0.11, 0.07] |
| Controllability | 0.00 | 0.04 | 0.02 | .987 | [−0.08, 0.08] |

**Note.** n = 1,219. The overall model was statistically significant, $F_{(8, 1210)} = 6.25$, $p < .001$, $R^2 = .040$, adjusted $R^2 = .033$. β = standardized regression coefficient; the reported SEs and 95% CIs correspond to the standardized coefficients. CI = confidence interval.

Changes in work strongly predicted longer AI use duration (β = 0.17, p < .001), consistent with a 'work intensification' interpretation: employees with heavier workloads are also those who have been integrating AI more into their routines (Jarrahi, 2018; Raisch & Krakowski, 2021). Skill variety also predicted duration (β = 0.13, p = .001), reinforcing its role as a facilitator of technology integration across multiple dimensions of use.

Task significance negatively predicted duration (β = -0.10, p = .022), indicating that employees in roles experienced as more consequential and impactful appear to use AI for shorter periods or integrate it less deeply. One interpretation is that high task significance may heighten concerns about delegating judgment to AI, leading employees to use it selectively rather than pervasively. Loss of status/position showed a marginally negative association (β = -0.09, p = .062), suggesting that professional identity concerns may discourage sustained AI engagement even when they do not block initial adoption.

## Discussion

These findings suggest that workplace AI adoption and use are shaped by *both* the structural features of jobs and employees' psychological appraisals of what AI implies for their work and professional standing. Among job characteristics, skill variety and autonomy were the most consistent predictors of whether employees reported adopting AI. This pattern aligns with work design theory, which argues that complex, varied work increases opportunities for learning and tool experimentation, and that autonomy provides discretion to trial and embed new practices (Hackman & Oldham, 1976; Morgeson & Humphrey, 2006; Parker & Grote, 2022). Importantly, these associations emerged even while accounting for multiple threat dimensions, suggesting that job design captures "opportunity structure" for AI use that is not reducible to general positivity (or anxiety) about AI.

At the same time, AI threat did not behave as a single, uniformly inhibitory construct. Instead, different threat dimensions related differently to outcomes, underscoring the value of treating threat as multidimensional (Brougham & Haar, 2018; Bankins et al., 2023; Trenerry et al., 2021). Perceived changes in work—capturing workload expansion and shifting expectations—was positively associated with frequency and duration. Given the cross-sectional design, this should not be interpreted as evidence that work intensification *causes* deeper AI integration; it may also reflect that employees who use AI more extensively are more exposed to workflow change and shifting demands. Still, the pattern is consistent with the possibility that AI becomes a tool employees turn to when they are navigating intensified work demands (Green, 2004; Kelliher & Anderson, 2010). In contrast, perceived loss of status/position showed a consistent negative relationship with depth of use, suggesting that identity-relevant concerns may be linked to more selective, bounded engagement even after adoption (Mirbabaie et al., 2021; Petriglieri et al., 2019). Finally, perceived loss of controllability showed a more mixed pattern, positively associated with adoption but directionally negative for depth of use, consistent with the idea that some employees may adopt AI in contexts where use is encouraged or expected, while still experiencing reduced control and consequently limiting how deeply they integrate it into everyday work (Parker & Grote, 2022).

Taken together, the results are consistent with a staged account in which the drivers of adoption differ from the drivers of depth of use. Adoption may be most closely tied to structural job affordances (discretion and task variety) that enable employees to experiment with AI and identify use cases (Hackman & Oldham, 1976; Morgeson & Humphrey, 2006). In contrast, deeper integration may hinge more on ongoing appraisals about how AI changes work demands, control, and role value (Bhattacherjee, 2001; Jasperson et al., 2005). This distinction echoes prior work showing that post-adoption behavior is not merely "more of" adoption; it often involves different mechanisms, constraints, and sources of resistance (Bhattacherjee, 2001; Ortiz de Guinea & Webster, 2013). In practice, employees can try AI without fully incorporating it into core workflows, especially when AI use raises concerns about professional identity or autonomy.

### Implications for theory

Three theoretical implications follow. First, the results reinforce calls to integrate work design into technology adoption models, which have historically emphasized beliefs and intentions while paying less attention to job structure and work systems (Davis, 1989; Venkatesh et al., 2003). Second, the findings support a multidimensional view of AI threat and suggest that treating "fear of AI" as unitary may obscure important distinctions—particularly between threat dimensions that may track exposure to change (e.g., work changes) and dimensions that may be more clearly inhibiting sustained engagement (e.g., status/position threat) (Bankins et al., 2023; Trenerry et al., 2021; Mirbabaie et al., 2021). Third, the pattern of results provides tentative support for theorizing adoption and depth of use as meaningfully distinct outcomes, consistent with continuance and post-adoption perspectives that separate initial uptake from longer-run integration (Bhattacherjee, 2001; Jasperson et al., 2005).

***Practical implications***

For organizations, the most actionable takeaway is that AI engagement is not simply a messaging or training problem. If adoption is strongly tied to skill variety and autonomy, then interventions that target only beliefs (e.g., "AI is helpful") may underperform when jobs do not provide discretionary space or recurring use cases. In those contexts, leaders may need to redesign workflows to create clear opportunities to apply AI, provide discretion over how it is used, and establish psychologically safe norms for experimentation (Parker & Grote, 2022; Zhang & Parker, 2019).

The association between perceived work changes and deeper AI use also warrants attention because it may reflect a "productivity pressure" dynamic in digital work: as tools increase capacity, expectations can expand, compressing timelines and increasing workload rather than freeing time (Mazmanian et al., 2013; Green, 2004; Kelliher & Anderson, 2010). Organizations can respond by monitoring workload and role creep during AI rollouts, clarifying what will *not* be added simply because AI is available, and aligning performance expectations with realistic AI-supported workflows.

Finally, although loss of status/position was not a statistically significant predictor of usage depth, its coefficients were consistently negative (e.g., for duration, $\beta = -0.09$, $p = .062$), suggesting a potential pattern in which identity-relevant concerns may be associated with more bounded or selective engagement after adoption. Addressing these concerns likely requires more than reassurance; it may involve redefining valued competencies, clarifying how human expertise remains essential, and communicating credible pathways for skill development and role progression in an AI-enabled workplace (Petriglieri et al., 2019; Mirbabaie et al., 2021; Trenerry et al., 2021).

***Limitations and future directions***

Several limitations constrain interpretation. First, the cross-sectional design precludes causal inference and makes temporal ordering ambiguous. For example, greater AI use could heighten perceptions of work change, but heightened work change could also motivate greater AI use. Longitudinal designs, especially those spanning pre- and post-deployment periods, would help clarify directionality and identify whether job design predicts adoption trajectories or whether adoption itself reshapes job characteristics and threat appraisals (Ployhart & Vandenberg, 2010). Experience-sampling approaches could further distinguish short-term within-person dynamics (e.g., daily workload spikes and AI use).

Second, the study relies on self-reported AI adoption and usage, which may be influenced by recall error or social desirability. Where feasible, future work should triangulate with objective indicators (e.g., system logs, frequency/duration telemetry, or tool-specific usage traces) and test whether self-report patterns replicate.

Third, the data come from a single organization, which limits generalizability. AI tools differ in capability, governance, and risk profile; implementation strategies differ in mandates versus voluntariness; and norms around appropriate use vary by region, function, and professional role. Replication across organizations and across types of AI tools (e.g., copilots vs. chat interfaces vs. workflow automation systems) would strengthen external validity.

Fourth, the models estimate average effects and may obscure heterogeneity. Future research should test theoretically motivated interactions and boundary conditions. For example, task significance may moderate the relationship between skill variety and AI depth of use, such that high-stakes roles show greater constraints on sustained use even when tasks are complex. Autonomy may buffer the negative association between identity-related threat and sustained use by preserving discretion in how AI is incorporated (Parker & Grote, 2022). At the implementation level, organizational climate and rollout practices may moderate these relationships (Klein & Sorra, 1996).

Finally, this study focused on adoption and depth of use, but not on *how* employees use AI (augmentation vs. substitution) or on downstream outcomes such as performance, well-being, learning, or skill development. Future research should examine whether deeper use is adaptive (e.g., improving performance or reducing strain) or potentially maladaptive (e.g., amplifying workload expectations or undermining learning), and for whom (Autor, 2015; Trenerry et al., 2021).

**Conclusion**

Workplace AI engagement appears to reflect an interplay between job structure and employees' appraisals of what AI implies for their work and identity. In this dataset, skill variety and autonomy were robust correlates of adoption, while threat dimensions showed differentiated associations with depth of use, particularly positive covariation between perceived work changes and deeper use and negative associations between status/position threat and sustained engagement. These findings should be interpreted cautiously given the cross-sectional and self-report nature of the data, but they underscore a practical point: building sustainable AI use likely requires more than training and encouragement. It may require aligning AI deployment with work design, monitoring unintended workload expansion, and proactively addressing identity-relevant concerns so that AI can be integrated in ways that support both performance and job quality.